# A Review of Machine Learning Techniques for Applied Eye Fundus and Tongue Digital Image Processing with Diabetes Management System

Wei Xiang Lim
Ph.D. Candidate
hcxwl1@nottingham.edu.my

Dr. Zhiyuan Chen
Assistant Professor
Zhiyuan.Chen@nottingham.edu.my

Dr. Amr Ahmed
Associate Professor
Amr.Ahmed@nottingham.edu.my

Dr. Tissa Chandesa
Research Training Development Manager
Tissa.Chandesa@nottingham.edu.my

Dr. Iman Liao
Associate Professor
Iman.Liao@nottingham.edu.my

## ABSTRACT

Diabetes is a global epidemic and it is increasing at an alarming rate. The International Diabetes Federation (IDF) projected that the total number of people with diabetes globally may increase by 48%, from 425 million (year 2017) to 629 million (year 2045). Moreover, diabetes had caused millions of deaths and the number is increasing drastically. Therefore, this paper addresses the background of diabetes and its complications. In addition, this paper investigates innovative applications and past researches in the areas of diabetes management system with applied eye fundus and tongue digital images. Different types of existing applied eye fundus and tongue digital image processing with diabetes management systems in the market and state-of-the-art machine learning techniques from previous literature have been reviewed. The implication of this paper is to have an overview in diabetic research and what new machine learning techniques can be proposed in solving this global epidemic.

## 1. INTRODUCTION

Diabetes Mellitus (DM) or diabetes is a long-term chronic disease, where the human body loses the ability to produce or respond to the hormone insulin. Due to this deficiency, abnormal metabolism of carbohydrates, fat, and proteins occur in the body. According to the World Health Organization (WHO) [1], 422 million adults estimated were living with diabetes in 2014. The global prevalence of diabetes has nearly doubled since 1980, rising from 4.7% to 8.5% in the adult population. In addition, data scientists had predicted that the prevalence of diabetes increases drastically in the year 2045. A report from the International Diabetes Federation (IDF) [2] projected that the total number of people with diabetes globally increase to 48%, from 425 million people (year 2017) to 629 million people (year 2045).

To reduce the prevalence of diabetes, existing innovative applications had been commercialised in the market today. Diabetes management system (DMS) is an innovative tool that assists diabetic patients to self-check their blood glucose level, calorie intake, and insulin doses. Traditionally, invasive DMS uses a blood glucose meter, a lancing device, and an online data management platform to keep track the blood glucose level. Conversely, non-invasive DMS had invented to ease patients. Patients do not require finger pricking themselves to keep track their blood glucose level. They only need to attach a small device onto the skin and the device will automatically keep track the blood glucose level in real-time without interfering their daily activities.

Until today, although there are many invasive and non-invasive DMSs (e.g. Accu-Chek [3] and Dexcom [4]) available in the market, scientific literature also reported that new research had been conducted to innovate the existing DMS [14-18]. Instead of collecting blood samples (e.g. invasively or non-invasively), eye and tongue images become another focused area to predict the prevalence of diabetes or to prevent diabetes complications. Therefore, machine learning plays an important role in this research area. Gulshan et al. [5] applied machine learning technique to develop an algorithm that detect Diabetic Retinopathy (DR) and Diabetic Macular Edema (DME) by using retinal fundus images. The result of the study achieved high sensitivity and specificity of 96.1% and 93.9% respectively. Machine learning (ML) is sub-field of Artificial Intelligence (AI) that provides machines with the ability to learn and improve over time from experience without giving any explicit instructions. Thus, this paper investigates innovative applications and past literature in the areas of DMS with applied eye fundus and tongue digital images. Different types of existing applied eye fundus and tongue digital image processing with DMS and state-of-the-art machine learning techniques from previous literature have been reviewed.

## 2. Literature Review

### 2.1 Diabetes Complications

Diabetes Mellitus (DM) is one of the common endocrine disorder, affecting 200 million people worldwide; and it is estimated that DM cases are increasing dramatically in the upcoming years [6]. The IDF [2] projected that the prevalence of DM increases significantly in most countries. For example, the Middle-East and North Africa region estimated that there would be an increase of 110%, from 39 million people in the year 2017 to 82 million people with diabetes in the year 2045.

Without proper monitoring, different types of diabetes complications can occur in different parts of the body. Patients would have an increased risk of developing various life-threatening health problems. The most common complication is Diabetic Retinopathy (DR). DR is a common terminology for all retina disorders. DR can be divided into two main types which include non-proliferative and proliferative DR. Non-proliferative DR (NPDR) is the earliest stage of retinopathy. However, NPDR can progress into deeper stages (mild, moderate, and severe) as more blood vessels become blocked in the retina. More severe DR would be Proliferative DR (PDR). PDR happens when the retina starts to grow new blood vessels. Growing new blood vessels often bleed, which may block vision partially or completely. Both NPDR and PDR will have signs of different types of lesions in the retina. According to IDF [2], DR affects over one-third of the population globally and is the leading cause of vision loss in working-age adults.

Uncontrolled blood glucose will affect patients' oral health as well. However, the awareness of oral manifestations and complications of diabetes are lacking worldwide [7]. Diabetic patients may suffer orally from periodontal diseases (e.g. gingivitis) and salivary dysfunction (e.g. reduction of salivary production, changes in saliva decomposition, and taste dysfunction). Oral fungal and bacterial infections have also reported in diabetic patients. Moreover, oral mucosa lesions which include stomatitis, geographic tongue, benign migratory glossitis, fissured tongue, traumatic ulcer, lichen planus, lichenoid reaction and angular cheilitis may also present in the oral region [8-11].

## 2.2 Machine Learning Techniques

Different machine learning techniques had been adopted in diabetic research. This paper addresses the machine learning techniques and the use of eye fundus and tongue images as dataset to conduct innovative research.

### 2.2.1 Machine Learning using Eye Fundus Images

Medical imaging plays a central role in diagnosing and treating diseases, including DR. Retinal image classification becomes the attention among researchers in the field of computer vision, where it carries potential benefits which will enable personalised health care and will provide physicians high quality diagnosis/therapy.

Mahiba and Jayachandran [12] achieved high accuracy in classifying glaucoma using retina fundus images. A total of 550 retinal images were used. The images were gathered using ZEISS retina camera (FF450) at the Government medical college in India. Convolutional Neural Network (CNN) and Support Vector Machine (SVM) were adopted in the classification. Their proposed model claimed to achieve an accuracy of 98.71%.

Another novel approach was Samant and Argawal [13] attempted to use iris images, instead of retina fundus images; to evaluate the feasibility of diabetes diagnosis. A total of 180 features were extracted to quantify the broken tissue information of the iris. Then, these extracted features were broken down into three groups which were first order statistics, textural features, and wavelet features. 10-fold cross-validation technique had applied in the study. Moreover, 6 different classifiers including Binary Tree Model (BT), Support Vector Machine (SVM), Adaptive Boosting Model (AB), Generalized Linear Model (GL), Neural Network (NN), and Random Forest (RF) had been used. Using different feature selections and classification methods were aiming to investigate the best feature selection algorithms and classification given the available dataset. Samant and Argawal [13] reported t-test feature selection yielded the highest classification accuracy almost in all classifiers. Among the different types of feature selections and classifiers, t-test feature selection and RF classifier performed the best with 89.63% of accuracy.

Furthermore, research has been conducted in the field of mobile computing and ophthalmology. Recently, Tan et al. [14] conducted a study that focused on Age-related Macular Degeneration (AMD). AMD is a form of eye disease that affects the elderly and diabetic patients. Tan et al. [14] developed a 14 layers CNN model to automatically detect and diagnose AMD accurately. Data was collected from Kasturba Medical Hospital in India. A total of 402 normal eye fundus, 583 retinal images with early, intermediate AMD, and 125 retinal images with wet AMD. Using the blindfold and 10-fold cross validation strategies, the CNN model achieved 91.17% and 95.45% accuracy respectively. Tan et al. [14] claimed that their solution is cost effective and portable. The advantage of the CNN model does not require feature extraction, selection, and classification. Moreover, the CNN model can be installed in a cloud system. Tan et al. [14] also mentioned that their solution can financially replace the medical grade SR screening equipment such as the Optical Coherence Tomography.

In addition, Toy et al. [15] suggested that the use of portable smartphone-based telemedicine system is to improve access to screening, surveillance and treatment of DR. Toy et al. [15] conducted a research to use a smartphone as a screening tool to detect referral warranted diabetic eye disease. A total of 50 adult patients with 100 eyes participated in this research. The research also compares smartphone-based results with clinical assessment of diabetic eye disease by standard dilated examination. Firstly, all patients underwent clinical assessment on ophthalmic examination. Next, patients underwent smartphone-assisted acquisition of spectacle-corrected near visual acuity and anterior/posterior segment photography. The phone is attached with an adapter containing a macro lens and external light source. All the patients' eyes were dilated with 1 drop of each of 2.5% phenylephrine and 1% tropicamide after visual acuity measurement. The results show that smartphone visual acuity was successfully measured in all eyes. Furthermore, smartphone-acquired fundus photography demonstrated 91% sensitivity and 99% specificity to detect moderate non-proliferative and worse diabetic retinopathy. Overall, this research demonstrates the potential use of a smartphone with low-cost adapted and lenses to screed for referral-warranted diabetic eye disease.

### 2.2.2 Tongue Image Analysis in Diabetes Diagnosis

Aside from using retinal fundus images, tongue images have been used as well. There are several advancements in tongue image analysis over the past decade [16, 17]. Prior to diabetes diagnosis, Zhang et al. [16] used tongue images (tongue body and tongue coating as features) to diagnose diabetes. Machine learning algorithms such as Support Vector Machine (SVM), Principle Component Analysis (PCA), and Genetic Algorithm (GA) were utilised. SVM was used to examine the tongue features. PCA was used to reduce the dimension of the tongue features. The result showed that the rate of prediction was 77.83%. After parameters normalization of the tongue images, the accuracy increased to 78.77%. Moreover, GA was adopted for feature selection and it enhanced the accuracy rate of cross-validation from 72% to 83.06%.

Similarly, Zhang and Zhang [17] used 672 images to differentiate between healthy and diseased tongue images to diagnose DM. Images were collected from Traditional Chinese Medicine Hospital but the images were classified based on western medical practice. Decision Tree (DT) was used to classify five different types of tongue shapes whereas SVM was used to classify diseases for each tongue shape utilising 13 geometric features. Furthermore, Sequential Forward Selection (SFS) was used to optimise the number of features. The average accuracy of disease classification is 76.24%.

Zhang, Kumar and Zhang [18] used tongue images to detect DM and NPDR. A total of 34 tongue features (colour, texture, and geometry) were used. K-nearest neighbour (k-NN) and Support Vector Machine (SVM) were adopted to classify the tongue features. The result showed that both machine learning algorithm achieved the same average accuracy for all 34 features. The highest average accuracy is 66.26% using the tongue geometry feature. By utilising Sequential Forward Selection (SFS) to optimise the feature selection, the average accuracy of SVM and k-NN were 80.52% and 67.87% respectively.

## 3. Discussion

In general, the main feature of the DMS is to monitor patients' blood glucose level either using invasive or non-invasive approach. However, these systems lack monitoring other important body parts as well such as the retina. Therefore, this paper proposes an additional feature that extend the existing DMS's functionality.

The proposed feature captures an image of the retina and detects the presence DR by computationally extracting and classifying the DR lesions present in the retina. The proposed feature enables users to self-check their retina, which in turn creates an awareness of their retina condition and seek professional guidance if there exist any complications. To achieve this research idea, Deep Learning (DL) is proposed as the ML model to detect the DR. DL has achieved high confidence in identifying, localizing, and quantifying pathological features in retinal disease [19]. EyePACS and Messidor will be used as dataset to train the proposed ML model. EyePACS is a free DR image dataset that can be obtained at Kaggle website [20]. It contains 35,126 training images and 53,576 testing images. Messidor is a publicly available dataset and it contains 1200 eye fundus colour numerical images [21]. Since a vast amount of dataset can be obtained easily, a DL model is proposed. The advantage of employing DL is due to its efficiency at handling large amount of data.

On the other hand, tongue image analysis can be an auxiliary diagnosis system within the primary DMS. Tania, Lwin and Hossain [22] stated that there is inconsistency in input images, for example, the image quality, image segmentation, and feature extraction. Therefore, the accuracy, robustness, and reliability are unconvincing. To address this problem, it is crucial that the obtained dataset must include, but not limited to tongue colour and texture, shape and geometry, oral lesions including stomatitis, geographic tongue, and fissured tongue to enhance the feature extraction. One possible solution is to adopt Transfer Learning (TL). TL is a ML technique that focuses on storing knowledge gained while solving one problem and applying it to a different but related problem. There is existing literature on tongue analysis [16-18] that had promising results that we can use and enhance the accuracy using TL.

## 4. Conclusion

In this study, an effort was made to identify and review the ML approaches applied on eye fundus and tongue digital image processing research, particularly in DR. To date, there are a number of significant studies carried out in the classification of DR using different ML techniques. Research ideas have proposed in this paper to extend the existing DMS by incorporating additional features to the device's functionality. Image processing is able to capture and analyse the captured retina and tongue images to detect DR and diagnose DM respectively. This proposed feature can create awareness of patients' retina condition and seek professional guidance if there exist such disease. Thus, it can minimise the risk of having latter stages of DR. Tongue image analysis can be an auxiliary diagnosis system to detect DM. Machine learning models, particularly Deep Learning, where it produces convincing result, is adopted in classifying DR. Furthermore, this paper addresses the challenge of adopting tongue image analysis due to the lack of quality dataset in this research area. Thus, Transfer Learning has proposed to solve the insufficient data and to enhance the result output.

## 5. ACKNOWLEDGMENTS